\documentstyle[12pt]{article}
\begin{document}
\begin{titlepage}
\rightline{\vbox{\halign{&#\hfil\cr
&\today\cr}}}
\vspace{0.5in}
\begin{center}
{\Large\bf
Phase Transitions and Mass Generation in 2+1 Dimensions}\\
\medskip
\vskip0.5in

\normalsize {G.W.Semenoff$^{\rm a}$, P. Suranyi$^{\rm b}$, and
L. C. R. Wijewardhana$^{\rm b}$}\smallskip
\medskip

{ \sl $^a$Department of Physics, University of British Columbia\\
Vancouver, British Columbia, Canada V6T 1Z1 and\\ $^b$Department of
Physics, University of Cincinnati\\ Cincinnati, Ohio, 45221
U.S.A.}\smallskip
\end{center}
\vskip1.0in

\begin{abstract}
The possibility that the epsilon expansion can predict the
order of phase transitions in three dimensional field theories is examined.
For a Hermitean matrix-valued order parameter, the epsilon expansion
predicts fluctuation induced first order phase transitions.
We analyze two 2+1-dimensional
quantum field theories which exhibit spontaneous symmetry breaking
and have martix order
parameters.  Using the large $N$ expansion,
we show that these models exhibit second order transitions and discuss
the implications for
the chiral symmetry breaking transition in 2+1-dimensional QCD for a
critical number of quark flavors.

\end{abstract}

\end{titlepage}
\baselineskip=20pt
\section{Introduction}

In many of the applications of quantum field theory where phase
transitions play a role, the order of the transition is often an
important question.  For example, the generation of nonzero baryon
number in the early universe requires, besides interactions violating
$B$ and $T$, that the system is out of equilibrium during the process.
This happens, most naturally, in conjunction with supercooling and
reheating phenomena near a first order phase transition. On the
contrary, a second order chiral phase transition is needed in theories
of dynamical mass generation, and also in composite Higgs theories, to
ensure that the vast array of mass scales generated dynamically lie
well below the scale of interactions responsible for the relevant
dynamics~\cite{nambu}.  The nature of the chiral phase transition in
QCD will also affect results of impending heavy ion experiments.

There has been a substantial amount of work on understanding the
detailed nature of the chiral symmetry breaking phase transition both
in four-dimensional and three-dimensional QCD.  Though there are some
important differences, the latter, more recently studied case of
three-dimensional QCD shares many of the features of its physical,
four dimensional relative. The question of dynamical mass generation
can be framed in much the same way as in four dimensions and the
technical complexity of the problem is very similar.  Many of the
analytical methods which are used in four dimensions can also be used
in three dimensions.  This gives a nontrivial test of the methods in a
slightly different context.  They can furthermore be compared to
numerical data which, though still very incomplete (particularly with
dynamical fermions), should be easier to obtain in a lower dimension.

Lower dimensional gauge theory can also be of physical interest, as in
phenomenological models of certain condensed matter and statistical
systems.  The existence or non-existence of, and the detailed nature
of phase transitions is of great importance for those models.

A variety of methods have been used to investigate the nature of
dynamical mass generation in 3 dimensional gauge theories. One of the
most important is the solution of Schwinger-Dyson equations in the
quenched ladder approximation.  Another very important method is the
$\epsilon$-expansion of Wilson and Fisher~\cite{wilson} around $d=4$.
(We shall call this the ``$4-\epsilon$ expansion.)  Other methods
include large $N$ expansions, strong coupling expansions, and the
$\epsilon$ expansion about two dimensions applied to chiral Lagrangian
sigma models (which we shall call the $2+\epsilon$-expansion).

In this paper we would like to discuss some of the relationships
between these methods.  Indeed, in the current literature there is an
apparent contradiction between the analysis of the
$4-\epsilon$-expansion and the use of Schwinger-Dyson equations.  We
shall comment on some of the issues involved.

\section{Three-dimensional QCD}

Consider the model defined by the action
\begin{equation}
S=\int d^{3}x\left(\bar\psi
\gamma\cdot D\psi+{N_cN_f\over4e^2}{\rm tr}F_{\mu\nu}^2\right),
\label{action}
\end{equation}
with $N_c$ colors and $N_f$ flavors of quarks: $\psi^a_\alpha$,
$a=1\ldots N_c$ , $\alpha=1\ldots N_f$. The Dirac matrices are $2\times2$
and Hermitean and $D=\partial+iA$,
$F_{\mu\nu}=\partial_\mu A_\nu-\partial_\nu A_\mu+i[A_\mu,A_\nu]$.

There is no chiral symmetry in 3 dimensions. Nevertheless, mass terms
can violate parity and time reversal invariance. Parity transformation
is defined as $\psi(x)\rightarrow\sigma_2\psi(x')$, for the fermions,
and $A_\mu(x)\rightarrow P_\mu A_\mu(x')$ for gauge bosons.  Here
$x'_\mu=P_\mu x_\mu$, where $P_1=-P_2=P_3=1$. The mass term,
$\psi^\dagger\sigma_3\psi$ is obviously odd under parity, but
preserves $U(N_f)$ global symmetry.

It is known that parity cannot be spontaneously broken in a
vector--like gauge theory \cite{witten}.  For even $N_f$ one can have
a mass term that is parity invariant, but breaks the flavor symmetry
to $SU(N_f/2)\times SU(N_f/2)\times U(1)$.  Such a mass term is of the
form
\begin{equation} L_m=\bar\psi Q_{N_f}\psi,
\label{mass-term}
\end{equation}
where $Q_{N_f}$ is a diagonal matrix with a set of $N_f/2$ eigenvalues
of 1 and a set of $N_f/2$ eigenvalues of -1.  Then we can define a
parity operator which exchanges fermions pairwise between the two
sets.

Whether the flavor symmetry is broken is a dynamical question.  The
symmetry breaking should generate an expectation value of a mass
operator and the physical fermions of the resulting model would be
massive.  This question is complicated by the fact that, as in four
dimensional QCD, the model likely confines color charge.  It has been
speculated that if chiral symmetry is not broken, even this three
dimensional theory, in spite of its severe infrared behavior, would
not be confining - the screening of the strong long ranged attractive
interactions which would be necessary to restore chiral symmetry would
also be sufficient to relax confinement~\cite{appelquist2}.  As yet,
this heuristic picture has little support from rigorous analytical
calculations, but is consistent with all known analytic and numerical
results.

The question of whether there can be a phase transition between
chirally symmetric and non-symmetric phases and if there exists such a
phase transition, what is its order, has attracted wide interest.
First, we observe that this question is unconventional in the sense
that we are not considering the model (\ref{action}) at finite
temperature, as would be the case in ordinary critical phenomena.
Instead, we are studying how the properties of the ground state of
(\ref{action}) change as we vary the relevant parameters, i.e. the
so-called quantum critical phenomena.  Second, we observe that, in the
conventional sense, the action (\ref{action}) does not contain any
coupling constants which can be varied in order to produce this phase
transition.  The QCD coupling, $e^2$, is dimensional and serves to cut
off the ultraviolet divergences~\cite{footnote1}.  Its presence
renders the model super-renormalizable.  We therefore do not expect
any phase transition as we vary $e^2$ - it is simply a dimensional
parameter which fixes the units by which we measure energy and
distance.  The only parameters which can be changed are the numbers of
colors and flavors, $N_c$ and $N_f$ respectively.  We ask the
question, does the ground state of (\ref{action}) exhibit different
phases for different $N_c$ and $N_f$?  Furthermore, if we vary these
continuously (which of course cannot be done in the physical world
since they are integers), what is the order of the phase transition?

\section{Review of investigations of symmetry breaking}

In this Section, we shall review some of the approaches to the
dynamical symmetry breaking problem for three dimensional QCD.
Generally, they are based on three different limits of the theory:
\begin{itemize}
\item the limit where $N_c>>N_f$ where planar diagrams dominate

\item the limit $N_f>>N_c$ where the bubble diagrams of the large
$N_f$ expansion dominate

\item the conventional lattice strong coupling limit $e^2/
\Lambda\rightarrow\infty$ where $\Lambda$ is the ultraviolet
cutoff

\end{itemize}

The most thoroughly analyzed limit of this model is that with
$N_f>>N_c$.  In that limit, internal gluon exchanges are suppressed by
factors of $N_c/N_f$ and the gauge interactions are very similar to
those in the Abelian three dimensional electrodynamics.  There have
been extensive studies of the symmetry breaking problem in the $1/N_f$
expansion using the self consistent Schwinger-Dyson
equations~\cite{appelquist2}~\cite{appelquist}.
  In the leading order, this approximation
sums both bubble and rainbow diagrams and obtains~\cite{appelquist2}
{}~\cite{appelquist}
the following picture: At large $N_f$ the symmetry is unbroken if
$N_f>N_f^{\rm crit}$, where $N_f^{\rm crit}=(64/3\pi^2)(N_c^2-1)/N_c$
and is broken if $N_f<N_f^{\rm crit}$.  The phase transition at
$N_f^{\rm crit}$ is of second order.  $N_f^{crit}$ is large enough
that the large $N_f$ approximation is thought to be quite accurate in
that region and corrections to $N_f^{\rm crit}$, for example, have
been shown to be small in the next-to-leading order.

The limit $N_c>>N_f$ has seen much less analysis.  In this limit, all
of the usual planar diagrams of the conventional large $N_c$ expansion
of QCD contribute to any process.  This limit is therefore more likely
to exhibit the features of a strongly interacting gauge field theory -
chiral symmetry breaking and confinement.

If the latter limit of the theory is indeed chirally non-symmetric and
confining, it could be investigated by effective lagrangian methods.
The quantum fluctuations in the broken symmetry phase are described by
the Grassmanian $SU(N_f)/SU(N_f/2)\times SU(N_f/2)$.  It can be
parameterized as the set of $N_f\times N_f$ hermitean matrix-valued
fields $\phi(x)$ which have $N_f/2$ eigenvalues $1$ and $N_f/2$
eigenvalues $-1$.  Algebraically, this can be expressed as the
condition that $\phi^2(x)=1$.  The effective action is the Grassmanian
sigma-model \cite{raj}
\begin{equation}
S_{\rm eff}=\int d^3x \frac{1}{2F}{\rm tr}
\partial\phi\cdot\partial\phi +{N_C}{4\pi}\int_{B_4} d^4x{\rm tr} \phi
(d\phi)^4
\label{sigmamodel}
\end{equation}
where we have added the final, topological term and where $B_4$ is a
four dimensional space whose boundary is the three dimensional space
on which the first term is defined.  As has been pointed out by
Feretti and Rajeev \cite{raj}, the presence of the topological term is
necessary to break an extra, unwanted symmetry of the first term
$\phi(x)\rightarrow -\phi(x)$ which is not a symmetry of QCD.
Furthermore, it is also necessary in order to obtain the correct
current algebra.  We observe that the number of colors, $N_c$ can
affect this model only through the dependence on the pion coupling
constant $F$ on $N_c$ and the coefficient of the topological term.

This model is difficult to analyze directly.  It is not renormalizable
and a large $N_f$ expansion corresponds to the as yet intractable
summation of planar fat graphs.  However, the coupling constant $F$ is
naturally of order $N_f/N_c$, so if the number of colors is much
larger than the number of flavors (in order to suppress the planar
graphs with many internal $\phi$-lines), this model should be
described by the weak coupling limit.  It has recently been studied in
the saddle point approximation \cite{soda} with results suggesting the
presence of a second order phase transition to the symmetric phase at
some critical value of $N_f/N_c$.

Furthermore, the model without the topological term has been studied
in the $2+\epsilon $ expansion by Brezin, Hikami and Zinn-Justin
\cite{brezin}.  They showed that it has an ultraviolet stable fixed
point at a non-zero value of the coupling constant $F$.  In ref.
\cite{soda} this was interpreted as implying a critical value of
$N_c/N_f$
above which the chiral symmetry is broken and below which the chiral
symmetry is restored.

These results are compatible with the general picture of symmetry
breaking which we advocate. However, as we shall discuss later, they
may not give a complete description of QCD in the present context in
that they neglect the presence of massless fermions at the phase
transition point.  We shall return to this point later.

The phases of three dimensional QCD have also been studied in the
strong coupling limit of lattice gauge theories~\cite{sodasemenoff}. It
was found that the strongly coupled lattice gauge theory is equivalent
to a $SU(N_f/2)$ Heisenberg antiferromagnet with spins taking values
in the Lie algebra~\cite{footnote2}
\begin{equation}
H_{\rm AFM}=\sum_{\rm lattice} S^a(x)S^a(y)
\end{equation}
where $x$ and $y$ are nearest neighbors on a bipartite lattice,
$a=1,\ldots,(N_f/2)^2-1$ and
\begin{equation}
\left[ S^a(x),S^b(y)\right]=if^{abc}S^c(x)\delta(x-y)
\end{equation}
is the Lie Algebra of $SU(N_f/2)$. The spins are in the representation
given by the rectangular Young tableau with $N_f/4$ rows and $N_c$
columns.

Furthermore, it was argued that the N\'eel order parameter of the
antiferromagnet is equivalent to the mass parameter of the lattice
gauge theory.  Large $N_c$ is the classical limit where the ground
state has N\'eel order.  The small $N_c/N_f$ limit is the quantum
limit for the antiferromagnet and is though to be disordered, at least
in dimensions less than four.  However the precise nature of the
ground state in that case is not known.  Between these limits there is
a phase transition which separates the ordered and disordered phases.
This transition is believed to occur at a critical line $N_f\simeq
N_c$ and to be of second order.  It has been conjectured that this
phase transition coincides with the chiral symmetry breaking
transition in three dimensional QCD \cite{soda}.

There are numerical studies of $U(1)$ gauge theory with $N_f$
flavors~\cite{kogut}. If the phase transition is indeed independent of
the color group, these investigations are also relevant.  They also
provide a critical value for the number of flavors.

\section{Effective bosonic theory and $4-\epsilon$-expansion}

Pisarski and Wilczek~\cite{pisarskiwilczek} and Wilczek~\cite{wilczek}
used analogies with condensed matter systems and the
$4-\epsilon$-expansion around dimension 4 of Wilson and
Fisher~\cite{wilson} to argue that the finite temperature chiral phase
transition of massless four dimensional QCD is second order for two
flavors
and first order for more than two flavors of quarks, irrespective of
the color gauge group. The first order transition is fluctuation
driven and it changes the simple mean field theory result that one
obtains using the Landau-Ginsberg effective action. In the literature
this phenomenon is called the Coleman-Weinberg
instability~\cite{coleman}.

Pisarski has further argued~\cite{pisarski} that a similar situation
arises in zero temperature three dimensional QCD. In the following we
shall review his reasoning.

First of all, it is assumed that the relevant fluctuating degrees of
freedom are bilinear combinations of fermion fields averaged over
color. They will be denoted by $\chi_{\alpha\beta}(x)$, defined as
\begin{equation}
\chi_{\alpha\beta}(x)=\sum_{a=1}^{N_c}
\langle\bar\psi^a_\beta(x)\psi^a_\alpha(x)\rangle.
\label{chi}
\end{equation}

The effective Landau-Ginzburg action for such bilinears in
$4-\epsilon$ dimensions is
\begin{equation}
S_{\rm LG}=\int d^{4-\epsilon}x\left\{\frac{1}{2}{\rm
Tr}(\nabla\chi\cdot\nabla\chi) +\frac{8\pi^2\mu^\epsilon}{4!}
\left[g_1({\rm
Tr}\chi^2)^2 +\frac{g_2}{4!}{\rm Tr}\chi^4\right]\right\}
\label{lg}
\end{equation}
Note that the effective theory is independent of $N_c$. Though the trace
of the
right hand side of (\ref{chi}) does not vanish boson $\chi$ in (\ref{lg})
is defined to be traceless.
   Based on general arguments, the symmetry breaking pattern
is such that parity is conserved.~\cite{witten} That is automatically
satisfied for a boson in adjoint representation, but not for the
$SU(N_f)$ scalar. Consequently, the $SU(N_f)$ scalar does not participate
in the critical dynamics.

Pisarski~\cite{pisarski} used $4-\epsilon $ expansion to investigate the
phases of the effective theory.  The beta functions of the effective
theory are
\begin{eqnarray}
\beta_1&=&-\epsilon g_1+\frac{N_f^2+7}{3}g_1^2+
2\frac{2N_f^2-3}{3N_f}g_1g_2+\frac{N_f^2+3}{N_f^2}g_2^3
\nonumber\\
\beta_2&=&-\epsilon g_2 +
2g_1g_2+2\frac{N_f^2-9}{3N_f}g_2^2.
\label{beta}
\end{eqnarray}
There is an ultraviolet stable fixed point at the origin.  There are
infrared stable fixed points in the perturbative regime only when
$N_f=2 $. (When $N_f<\sqrt{5}\sim 2.3$ there is an infrared stable
fixed point at $g^*=(6\epsilon/(N_f^2+7),0)$ where the model
(\ref{lg}) is in the same universality class as the $O(N_f^2-1)$
non-linear sigma model.)  When $N_f\geq3$ there are no infrared stable
fixed points.  The situation is very similar to the one encountered by
Coleman and Weinberg~\cite{coleman}. There, the existence of two
competing coupling constants allowed the existence of a minimum of the
effective potential in the perturbative regime, leading to a
fluctuation induced first order phase transition. Conditions for the
existence of first order phase transitions can be obtained from the
$\beta$-functions alone as it has been shown a long time
ago~\cite{yamagishi}. The condition, beside the absence of infrared
stable fixed points, is that the renormalization group trajectories
cross the stability line,
\[
4(g_1+g_2)+\beta_1(g_1,g_2)+\beta_2(g_1,g_2)=0,
\]
 a line at which the potential is minimized. Provided the potential
vanishes at zero field, this should happen in the region where the
potential is negative, $g_1+g_2<0$. For $N\geq3$ the trajectories
generated by the beta functions of (\ref{beta}) do cross the line of
stability. The conclusion of Pisarski is that there is a first order
phase transition for $N_f\geq3$, and parity conserving mass is
generated spontaneously for all $N_f$.

This result seems incompatible with the picture of a chiral phase
transition at some critical value of $N_c\sim N_f$ which we have
described in the previous sections.  In the remainder of this Section
we shall list some of the possible reasons for this apparent lack of
agreement.

\begin{itemize}

\item In the conventional applications of the $4-\epsilon$-expansion to
study  the critical behavior of a finite temperature 4 dimensional
theory, the fermions cannot contribute to critical fluctuations because
they have antisymmetric boundary conditions in the time direction and
thus their `mass' is $O(1/\beta)$.  Therefore, fermionic degrees of
freedom need not be taken into account in the computation of the
$\beta$-function.  In the zero temperature 3 dimensional theories which
we are studying here, however, the mass of fermions vanishes at the
chiral transition point, thus their contribution is
 important and
should also be taken into account. Thus if the theory has only four fermi
 interactions
and no gauge interactions it belongs to the universality class of
Yukawa theories. Therefore to do a reliable $\epsilon$ expansion analysis one
should  add an appropriate Yukawa coupling
to the scalar theory in equation (7) and look for fixed points of the
resulting beta functions.
In the case of QCD3 the situation is somewhat complicated.
If confinement effects are small , one must also include
gauge interactions in the computation of the beta functions.
It has been conjectured  that when the number of
fermion flavors  exceeds a critical value ,the long range attractive
 gauge interactions
get screened leading not only to chiral symmetry restoration but also to
deconfinement ~cite{appelquist2}.
If this is the case Pisarski's calculation should be appropriately modified
to take into
account the added complications. A calculation  of effective scalar potential
for QCD3 including gauge and fermion contributions has been carried out in
the large flavor limit~\cite{ATW}. It is seen  that the effective
potential shows  non analytic behavior
as the critical point is approached from the symmetric phase.

\item The symmetries of the effective action (\ref{lg}) are not
quite the same as that of three dimensional QCD described by
(\ref{action}).  Action (\ref{lg}) has a reflection symmetry,
$\chi\rightarrow-\chi$, which is distinct from parity and which is
absent from the original gauge theory. This symmetry cannot be broken
by relevant terms in the framework of the Landau-Ginzburg approach.
It is broken by the topological term in the sigma model
(\ref{sigmamodel}).  However, it is impossible to take into account
the influence of the topological term in the epsilon expansion since
it cannot be continued in dimensions.

\item There are known examples of non-perturbative infrared stable
fixed points. Even in the scenario of the $4-\epsilon$-expansion
starting with a Landau-Ginzburg effective action, that is valid in the
large $N_c$ limit, it is very possible that there exists a
non-perturbative infrared stable fixed point.  Such fixed points are
known to occur in 3-dimensional models with fermions and four-fermion
interactions
\cite{wilson,rosenstein}.  For example,
consider~\cite{wijewardhana}
\begin{eqnarray}
S=\int d^{3}x\left(\bar\psi(
\gamma\cdot
\partial+\phi)\psi +
\frac{N\Lambda}{2\lambda}
\phi^2\right)
\label{action1}
\end{eqnarray}
Eliminating $\phi$ by using its equation of motion, produces the
4-fermi coupling $-\frac{\lambda}{2N}(\bar\psi
\psi)^2$, which is invariant under a global $SU(N)$ flavor symmetry.
This model is renormalizable in the large $N$ expansion.  Integrating
the Fermions gives the effective scalar field theory
\begin{equation}
S_{\rm eff}=-N{\rm Tr}\ln(\gamma\cdot
\partial+\phi)+
\frac{N\Lambda}{2\lambda}\int\phi^2
\end{equation}
for which $1/N$ is the coupling constant.  It is known that this model
has a non-perturbative fixed point at large $\lambda$.  Note the
difference between this model and gauge theory.  Since the former is
strictly renormalizable, the ultraviolet cut-off $\Lambda$ appears as
a dimensionful parameter in addition to the dimensionful coupling
$\lambda$, thus providing an additional parameter $\lambda\Lambda$.
It is this parameter which is tuned to obtain critical behavior.

\item The $4-\epsilon$ expansion is known not to be reliable in some
models.
The Higgs-gauge model with $N$ complex scalars, which is a model for
superconductors, was found by Halperin, Lubensky, and
Ma~\cite{halperin} to have a first order phase transition, due to the
lack of infrared stable fixed points. This is a generalization of the
Coleman-Weinberg argument. It is known that this system in the Type II
regime (roughly speaking the Higgs mass is larger than the gauge mass)
has the same critical properties as the smectic-nematic phase
transition in liquid crystals.  Experimentally, the latter system was
found to have a second order transition with $XY$ exponents.  The
problem has been studied by Monte Carlo simulations~\cite{halperin2}.
It was shown that at $\epsilon=1$ there is a new fixed point with $XY$
exponents, which is absent at $\epsilon=0$. March-Russell~\cite{marc}
came to the same conclusion by studying the gauge interactions of $N$
complex $p$-vector Higgs bosons in $U(p)$-gauge theory and comparing
$4-\epsilon$ and $2+\epsilon$ expansions.  In general, sigma models
have been studied in the $2+\epsilon$ expansion by Brezin, Hikami, and
Zinn-Justin. They found fixed points and nonzero coupling for
$\epsilon>0$. This is in contradiction with results obtained when
one investigates linear sigma models in $4-\epsilon$ expansion, though
the physical degrees of freedom driving the phase transition and the
universality class are believed to coincide in the two models.

\end{itemize}

If, as we have noted above is possible, the analysis of the
$4-\epsilon$ expansion is reliable in the limit of large $N_c$, it is
still implies that the chiral phase transition is a fluctuation
induced first order transition. This comes about by the following
reasoning: In the zero temperature gauge theory defined by action
(\ref{action}), the parameters are $N_f$, and $N_c$.

The coefficients
of the operators in the effective Landau-Ginzburg theory (\ref{lg}) may depend
on the parameters of the model.  In the case of QCD in 2+1 dimensions, the
only relevant parameters are $N_c$ and $N_f$.  There is a phase transition
if it
of possible to tune $N_c$ and $N_f$ so that the quadratic terms in the
Landau-Ginsberg potential vanish. This would give a critical curve in the
$N_c-N_f$--plane.   The results of ~\cite{appelquist} imply that this curve is
approximately linear, $N_f\simeq N_c$. The question of whether the
transition is
second order or a fluctuation induced first order transition remains.
It is not clear that the  analyses of the phase transition
described in Section 3  distinguish a second order transition
from one which is weakly first order.  This could in principle
be resolved using an improved $4-\epsilon$ expansion.  Also, the model which
we construct in the next Section has the same flavor symmetries as
2+1-dimensional QCD and has a second order transition.  If they indeed
lie in the
same universality class, this would indicate that the chiral
transition at $N_f\sim N_c$ in QCD is indeed second order.

\section{A gauge model with global symmetry $U(N_f)\otimes Z_2^{\rm parity}$}

In this Section we shall construct a model which has the same global
symmetries as QCD, is renormalizable in the large $N$
expansion, and has a second order chiral phase transition. Since the
model shares the Landau-Ginzburg potential with QCD, claims of a proof
of first order phase transition based on the Landau-Ginzburg potential
and the $4-\epsilon$-expansion are invalidated.
The model will consist of four-fermion interactions which drive the phase
transition.  To illustrate, we begin with a simpler version, a theory with
only four fermion interactions and where the color symmetry is not gauged.

\subsection{A toy model with 4-Fermion interaction}

Let us begin this section with a toy model with 4-fermion interactions
with $N_f$ ``flavors'' and $N_c$ ``colors'' of fermions
$\psi_a^\alpha$ $a=1\ldots N_c$, $\alpha=1\ldots N_f$ and an action
which is a generalization of (\ref{action1})
\begin{equation}
S=\int
d^{3}x\left(\bar\psi(
\gamma\cdot \partial+\phi)\psi
+\frac{N_c\Lambda}{2\lambda}{\rm tr}(\phi^2)\right).
\label{action2}
\end{equation}
Here $\phi_{\alpha\beta}$ is a traceless $N_f\times N_f$ Hermitian matrix.
Integrating over $\phi$ gives the four-fermion interaction
\begin{equation}
-\frac{\lambda}{2N_c\Lambda}
\int d^3x \bar\psi_aT_A\psi_a\bar\psi_bT_A\psi_b,
\label{four-fermi}
\end{equation}
where both flavor and color indices have been
omitted. $T_A$ is the generator of the flavor symmetry group, $SU(N_f)$.
One could modify interaction term (\ref{four-fermi}) by
including an isoscalar four-fermi interaction, $(\bar\psi\psi)^2$.
That case will be discussed
briefly later.

This model can be readily analyzed in the large $N_c$ expansion.  If the
color group were gauged, the necessity of summing planar diagrams would
render this expansion intractable.  In the following section we shall
suggest a way to overcome this difficulty.  In this section we shall
study the model (\ref{action2}).

The fermion fields can be integrated out to give in leading order of
$N_c$ the following effective potential for the scalar field
\begin{equation}
V[\bar\phi]= N_c{\rm\bf
tr}\left[\left(\frac{\Lambda}{\lambda}-\frac{\Lambda}{\pi^2}
\right)\frac{\bar\phi^2}{2}+\frac{1}{6\pi} [\bar\phi^2]^{3/2}\right]
\end{equation}
where the trace is over ``flavor'' indices.  This potential predicts a
second order phase transition at the point where the effective mass
parameter
\begin{equation}
t\equiv\frac{\Lambda}{\lambda}-\frac{\Lambda}{\pi^2}
\label{mass}
\end{equation}
vanishes. When $t<0$, the field $\phi$ acquires a vacuum expectation
value, breaking the $U(N_f)$ symmetry. The tree approximation gives
the mean field value, 0, for the exponent $\gamma$, defined by
\begin{equation}
V^{\rm crit}[\bar\phi]\sim {\rm\bf tr}\vert\bar\phi\vert^{d(1+\gamma)}.
\label{gammadef}
\end{equation}
As we can see there is still no dependence of the universality class
on $N_c$ at tree level.

The effect of higher order corrections will be sketched below.  In
$O(1/N_c)$ one has to calculate contributions coming from the
fluctuations of the fields. The corrected potential has the form
\begin{equation}
V[\bar\phi]=N_c\left[\frac{t}{2}{\rm\bf tr}\bar\phi^2+
\frac{1}{6\pi}{\rm\bf
tr}(\bar\phi^2)^{3/2}+\frac{1}{N_c}{\rm\bf TR}\ln\Delta\right],
\label{pot}
\end{equation}
where $\rm \bf TR$ is a trace over spacetime as well as flavor indices
and
\begin{equation}
\Delta_{ijkl}(x,y)=\frac{1}{\lambda}\left(\delta_{il}\delta_{jk}
-\frac{1}{N_f}\delta_{ij}\delta{kl}\right)
\delta(x-y)+{\rm Tr}(x\vert
\frac{1}{\gamma\cdot
\partial+\bar\phi}\vert y)_{il}(y\vert\frac{1}
{\gamma\cdot\partial+\bar\phi}\vert x)_{jk},
\end{equation}
is the self-energy of the scalar to order $N_c^{-1}$.

The global flavor symmetry can be used to diagonalize the order
parameter $\bar\phi$.  Here, we shall assume the symmetry breaking
pattern
(when $N_f$ is even) which preserves parity and an $U(N_f/2)\times
U(N_f/2)$ flavor symmetry
\begin{equation}
\bar\phi={\rm \bf diagonal}\left(m,m,\ldots,m,-m,-m,\ldots,-m\right).
\end{equation}
Then, the effective potential has the form\cite{footnote3}
\begin{equation}
V_{\rm eff}=
\frac{N_cN_f}{2}t m^2+\frac{N_cN_f}{6\pi}m^3+
\frac{2N_f^2}{3\pi^3}m^3\ln\frac{m}{\mu}
{}.
\label{effpot}
\end{equation}
The cubic term of the effective potential comes from the fermion ring
vacuum diagram, while the last term from the boson ring vacuum
diagram.  The effective mass parameter $t$ also renormalizes at this
order.  We have also cancelled a logarithmic ultraviolet divergence by
a wave-function renormalization for the scalar field, $m\rightarrow
zm$ and appropriate choice of $z$.

The logarithmic term in the effective potential is reminiscent of a
similar term in the Coleman-Weinberg \cite{coleman} effective
potential in four dimensional theories.  Furthermore, as in the
Coleman-Weinberg analysis, the critical model (at $t=0$) indeed has a
minimum away from the origin and an apparent first order phase
transition with the exponentially small order parameter
\begin{equation}
m_0=\mu\exp\left(-\frac{\pi^2}{4}\frac{N_c}{N_f}\right)
\end{equation}
However, as in the Coleman-Weinberg scenario with one coupling
constant, this minimum is at a point outside of the perturbative
regime - it is for an exponentially small order parameter and the
logarithm in the effective potential invalidates the perturbative
result in the region where the order parameter is close to zero.

As usual, the trouble with the effective potential near $m=0$ can be
fixed using the renormalization group.  This amounts to understanding
how higher orders in logarithms would contribute to the effective
potential.  For the present case (unlike in the four-dimensional model
considered by Coleman and Weinberg) we shall argue in the Appendix
that the corrections exponentiate to produce an effective potential of
the form
\begin{equation}
V_{\rm eff}=
\frac{N_cN_f}{2}tm^2+\frac{N_cN_f}{6\pi}m^{3[1+4N_f/(3\pi^2N_c)]}.
\label{expo}
\end{equation}
This improved effective potential exhibits a {\bf second order} phase
transition at $t=0$.

The exponent $\gamma$ defined in (\ref{gammadef}) is $O(1/N_c)$,
$\gamma=4N_f/(3\pi^2N_c)$. The exponent is identical with the
anomalous dimension of the boson.  In other words, the universality
class depends on $N_c$. In the broken phase, $\mu^2<0$, the
eigenvalues of matrix $\langle\phi\rangle$ are $m_i=\pm m_0$, where
$m_0\sim(|t|)^{1/(1+3\gamma)}$.

It is worth mentioning what happens if a flavor neutral term is
included in the
four-fermion interaction. Then in addition to the
adjoint representation scalar boson (isotensor boson), one is forced
to introduce a flavor neutral (isoscalar) boson as well. The
isotensor and isoscalar bosons have, in general, different couplings
to fermion pairs. On one hand, new divergent diagrams appear and both
couplings start to run. The anomalous dimension of bosons is not any
 more twice the anomalous dimension of fermions. On the other hand,
some crucial features of the pure isotensor theory survive. The
logarithmic derivatives of couplings and the anomalous dimensions
are independent of the couplings constants; they depend on $N_f$
and $N_c$ only. The running fermion-boson coupling constant
 exponentiate into forms $g\sim m^{\lambda(N_f)/N_c}$. In a similar
 manner, the three boson couplings exponentiate as well. Thus, even
in this case, the effective potential exponentiates into a pure $N_f$
and $N_c$ dependent power of the the expectation value of the field,
$m$. The phase transition is of second order and critical exponents
are dependent on $N_f$ and on $N_c$. Details of the discussion of
this model will be given elsewhere.

\subsection{Gauging the ``color'' group: Chern-Simons gluo-dynamics}

The four-fermi theory which we have been discussing has a
second order phase
transition.  However, the
flavor symmetries is very different from that of 2+1-dimensional QCD.
One can come
closer to QCD and remove the color degeneracy of the model by gauging
the color symmetry.  We shall introduce gauge fields in such a way as
to preserve, as much as possible, the solvability of the four-fermion
model in the large $N_c$ limit.

Consider the action
\begin{eqnarray}
S=\int d^3x \left( \bar\psi(\gamma\cdot D_A+\phi+M)\psi
+\bar\chi(\gamma\cdot D_B+\phi-M)\chi
\right.
\nonumber\\
\left.
+\frac{N_c\Lambda}{2\lambda}{\rm\bf tr}(\phi^2)+\frac{ N_c}{2g^2} {\rm\bf
tr}\left(AdA+\frac{2}{3}A^3-BdB-\frac{2}{3}B^3\right)\right)
\label{gauged}
\end{eqnarray}
where $D_A=\partial+iA$ and $D_B=\partial+iB$.  Here, we have assumed
that there are an even number of colors and have gauged the group
$SU(N_c/2)\times SU(N_c/2)$.  As a kinetic term for the gauge field, we
have introduced the Chern-Simons terms with a dimensionless coupling
constant $g$.  This makes the gluon non-propagating. It is very
important to observe that even though there are Chern-Simons terms
in the theory, it is still parity invariant. In fact, this theory has
the same flavor symmetries as QCD: it is parity symmetric, the parity
transformation, besides the obvious spacetime transformation, also
interchanges the two
types of fermions $\chi$ and $\psi$ and the two types of gauge fields
$A$ and $B$.  This symmetry insures that if there is no bare fermion mass
it will not be generated to any finite order in perturbation theory.

This theory also has the flavor symmetry $U(N_f)$ and one can study
the possibility of a phase transition with symmetry breaking pattern
$U(N_f)\rightarrow U(N_f/2)\times U(N_f/2)$.  Its one large
difference with QCD is that here the fermions have a bare mass, $M$.
They are not massless at the critical point unless we also tune $M=0$.
This model thus allows us to explore two possibilities, one which is
like QCD with $M=0$ and where massless fermions are important to the
critical behavior and the other where $M\neq 0$, fermions are all
massive in the region of $\phi\sim0$ and one would expect that perhaps
the Landau-Ginsburg theory and epsilon expansion gave an accurate
description of the fluctuating degrees of freedom.

For the gauge interactions, to obtain the large $N_c$ limit one must
sum all of the planar diagrams which contribute to the effective
potential.  The leading order is $N_c^2$ and contains no fermion
loops.  It therefore does not couple to the order parameter $\bar\phi$
and does not contribute to the effective potential for $\bar\phi$.  The
next to leading order, $\sim N_c$, contains one fermion loop, no
scalar lines and all insertions of planar gluon lines.  The problem of
summing all contributions of planar gluons is as yet intractable.
Here, we shall remedy this situation by using a simultaneous large
$N_c$ and small $g^2$ expansion~\cite{footnote4}. In this expansion,
unlike conventional gluo-dynamics where
the gluons would have a Yang-Mills action, the Chern-Simons
gluo-dynamics is infrared finite, i.e.  infrared divergences do not
ruin the perturbative expansion \cite{chen}.

Then, when $M=0$ (the QCD case), the order $g^2N_c^0$ contribution to
the effective potential vanishes.  Then, in this model, the effective
potential is given by (\ref{effpot}) with corrections being of order
$N_c^kg^{2p}$ with $k+p\geq 1$. Furthermore, since the gauge
interactions do not contribute to the logarithmic divergence to the
order we have considered, the renormalization group analysis as well
as the result (\ref{expo}) are also valid.  Thus, this theory has a
{\bf second order} phase transition and has the same flavor symmetry as QCD.

The analysis of this problem when $M\neq0$ is more difficult and thus
far is done only to the leading order.  There, the effective potential
has the form
\begin{equation}
V_{\rm eff}^{M}[\bar\phi]=N_c{\rm\bf tr}\left\{\matrix{ \frac{t}{2}M^2+
\frac{M^3}{6\pi}+(\frac{t}{2}+\frac{M}{2\pi})\bar\phi^2
&~ \vert\bar\phi\vert\leq M\cr
\frac{t}{2}M^2+\frac{M^2}{2\pi}\vert\bar\phi\vert+\frac{1}{6\pi}
\vert\bar\phi\vert^3
&~\vert\phi\vert\geq M\cr}\right\}
\label{Meffpot}
\end{equation}
This system has a marginally first order behavior.  When
$t>t_c=-M/2\pi$ the minimum is at $\bar\phi=0$.  Then at $t=t_c$ the
effective potential has zero curvature in the region
$\vert\bar\phi\vert\leq M$ and is convex where $\vert\bar\phi\vert>M$.
When $t>t_c$ the minimum begins at $\vert\bar\phi\vert=M$ and
increases as $t$ decreases.  This is very similar to a first order
behavior: the order parameter $\bar\phi$ jumps from $\bar\phi=0$ to
$\vert\bar\phi\vert=M$ at $t=t_c$.

It is intriguing that this ``almost first order'' behavior occurring for
this case is probably closer to the prediction of
the epsilon expansion than
the case of pure QCD.  However, even here we conjecture that higher
order corrections smooth the behavior of the effective potential and
the phase transition remains second order.
Detailed behavior of this model in the next order of
expansion are interesting and are the subject of further study.

\section{Conclusion}

In this Paper, we have analyzed a pure four-fermion model and a gauged
four-fermion model in the large $N_C$ limit.  We have shown that both of
these models exhibit second order phase transitions.

Perturbatively, the models appear to have fluctuation induced first order
transitions.  However, we have shown that higher order corrections
lead to exponentiation of logarithmic terms and restoration of second
order behavior.

In the gauged four-fermi model, we could consider two cases -- one where
the fermions are massless at the phase transition and the other where they
are massive.  In the case where they are massless, at least to the order of
perturbation
theory which we considered, the nature of the
phase transition was identical to that in the pure four-fermion theory.
In the case where the fermions are massive, however, the behavior was
 markedly
different -- it exhibited an ``almost first order'' phase transition.  It is
intriguing that the $4-\epsilon$ expansion predicts a fluctuation induced
first order transition for this model.  It is also interesting that as the
fermion bare mass goes to zero, the transition becomes second order (the jump
in the order paramater as the coupling is varied through the critical point
is proportional to the bare mass).

\appendix

\section{Appendix}
The exponentiation of the potential is proven by the method of
renormalization group. The field theory we consider is defined by
action (\ref{action2}). The fermion bubble corrections are summed to
give a modified boson propagator, which has the form
\begin{eqnarray}
D_{ij}^{-1}(p)&=&\frac{1}{2}\left(\frac{\Lambda}{\lambda}-
\frac{\Lambda}{\pi^2}
\right)+\frac{m}{2\pi}+
\frac{p^2+(m_i+m_j)^2}{4\pi p}\tan^{-1}\frac{p}{2m}
\nonumber\\&=&\frac{1}{2}\left(
\frac{\Lambda}{\lambda}-\frac{\Lambda}{\pi^2}
\right)+\frac{p}{8}+\frac{(m_i+m_j)^2}{8p}-
\frac{m[3(m_i+m_j)^2-4m^2]}{6\pi p^2}+O(m^4p^{-3}),
\label{boson-prop}
\end{eqnarray}
where $m_i=\pm m$ is the mass of the $i$th fermion.

The renormalization group functions can easily be calculated in
$O(N_f/N_c)$.  The only primitive divergent diagram is the fermion
self-energy correction diagram in this order.  Thus,
the coupling constant
does not run. The logarithmically divergent part of the
fermion self energy diagram is
\begin{equation}
\Delta S_F^{-1}(k)\simeq \frac{4N_f}{3N_c\pi^2}
\not\! k \ln\frac{\Lambda}{\mu}.
\label{prop2}
\end{equation}

The boson propagator gets logarithmic corrections only due to the
dissociation of the boson into two fermions. Thus, the leading order
anomalous dimension of the boson is twice the anomalous dimension of
the fermion.  In other words, as shown by (\ref{prop2}),
\begin{equation}
\gamma_b=\frac{4N_f}{3N_c\pi^2}.
\label{gamma}
\end{equation}

At the critical point, the
Callan-Symanzik equation
\begin{equation}
\left(\frac{\partial}{\partial t}+\gamma_bm
\frac{\partial}{\partial m}\right)V(m)=0
\label{callan}
\end{equation}
has the boundary condition
\begin{equation}
\lim_{N_f/N_c\rightarrow0} \frac{V_0}{N_cN_f}=\frac{1}{6\pi}m^3.
\label{boundary}
\end{equation}
Here $t=\log\mu/m$.
Then the solution of the Callan-Symanzik equation is straightforward.
\begin{equation}
V_{{\rm eff}}=\frac{1}{6\pi}m^3\exp\left\{-3 \int_0^t{\rm
d}t'\frac{\gamma_b(t')}{1-\gamma_b(t')}\right\}.
\label{final}
\end{equation}
Substituting (\ref{gamma}) into (\ref{final}) gives exponentiated form
(\ref{expo}).

\section*{ACKNOWLEDGEMENTS}

 The work of G. S. is supported by the Natural Science and Engineering
Research Council of Canada.  The work of P. S. and of L. C. R. W. is
supported in part by the United States Department of Energy under
grant no.  DE-FG02-84ER40153.  L.C.R.W. thanks T. Appelquist and J.
Terning for valuable discussions.


\begin{thebibliography}{10}
\bibitem{footnote1}  This is somewhat analogous
to four dimensional QCD where the dimensional transmutation which
accompanies asymptotic freedom changes the coupling constant into the
mass scale.  Both theories are weakly coupled at high energies.
\bibitem{footnote2} This is, strictly speaking, only true when
$N_f$ is a multiple of four.  The reduction of the flavor symmetry
group from $SU(N_f)$ to $SU(N_f/2)$ occurs by explicit symmetry
breaking by the lattice regularization.  In that case, the continuous
part of what we refer to as chiral symmetry is actually broken
explicitly.  However, a discrete chiral symmetry remains and is
sufficient to render fermions massless.  It is therefore reasonable to
ask whether fermion mass is generated in the lattice models.  This
mass generation is found to occur as N\'eel order of the effective
antiferromagnet.
\bibitem{footnote3} {The method for
computing the integral is outlined in ref. \cite{soda}.}
\bibitem{footnote4} One could alternatively view
this expansion as a large $N_c$ expansion where $g^2$ is also of order
$1/N_c$.
\end{thebibliography}
 \end{document}